\documentstyle [12pt,twoside]{article}
  
\begin{document}
\begin{flushright}
UBCTP--93--006
\end{flushright}
\vspace*{1cm}
\centerline{\large{\bf Almost Readily Detectable}}
\centerline{\large{\bf Time Delays from Gravity Waves ?}}
\vspace*{1cm}
\centerline{\bf Redouane Fakir}
\vspace*{0.5cm}
\centerline{\em Moroccan Center for Scientific Research}
\centerline{\em 52 Ave. Omar Ibn Khattab}
\centerline{\em BP 1364 Agdal, Rabat}
\centerline{\em  MOROCCO}
\vspace*{0.5cm}
\centerline{\em Cosmology Group, Department of Physics}
\centerline{\em University of British Columbia}
\centerline{\em 6224 Agriculture Road}
\centerline{\em Vancouver, B. C. V6T 1Z1}
\centerline{\em CANADA}
\vspace*{0.5cm}
\centerline{April 1993}
\vspace*{0.5cm}
\centerline{\bf Abstract}
\vspace*{0.5cm}

When a source of gravity waves is conveniently placed between
the Earth and some source of light, preferably a pulsating
source, the magnitude of time delays induced by the gravity
waves could, in optimal situations, be not too far out of the
reach of already existing technology. Besides the odd case of
near-to-perfect alignment one might be lucky enough to
encounter in the Galaxy, there exists several astronomical
sites where good alignment occurs naturally. A good example
is when the light source and the gravity-wave source
are, respectively, a high-frequency pulsar and a neutron
star, locked together in a tight binary. We are lead
to believe that neutron-star gravity waves might be directly
observable in timing data of systems such as PSR B1913+16.

\clearpage
\centerline{\bf I. Introduction}
\vspace*{0.5cm}
It is now well established that any direct, unequivocal detection
of gravity waves would have momentous consequences on both
observational astronomy and fundamental physics[1-19].
However, due (so we think,) to the extreme weakness of the
waves when they reach the Earth, no such detection has
yet occurred, despite almost thirty years of
vigorous efforts and considerable progress on both
the theoretical and the technological fronts. It is particularly
unfortunate that some of the most copious and most interesting
gravity waves, the low-frequency waves from astrophysical
sources such as binary stars, seem bound to remain
out of the reach of the few developed detectors, chiefly because
of seismic noise. It is hoped, nevertheless, that space-based
experiments will be able to detect low frequency waves
by the beginning of the next century.

Detection prospects are no brighter for the shorter
waves expected from neutron stars. The extreme weakness
of these puts them hopelessly out of detectability range
for most of the techniques envisaged so far.

Recently, however, the case was made that
there were ways, after all, of detecting
astrophysical gravity waves (such as those from binary
stars and neutron stars) sooner than
previously thought [20-24].
The basic idea
is that there are situations where one can
observe gravity waves where they
are still relatively strong, that is, very near the source.
If it is ultimately confirmed that this is indeed doable,
then the observational situation
could be suddenly improved by over
several orders of magnitude. Of course, even if
ultimately successful, this approach cannot replace the vigorous
and long-term effort necessary for detecting gravity waves in the
generic
case, where the waves can come from arbitrary directions and
with the usual extremely small amplitudes.

In [21], it was shown that a gravitational
``pulse with memory of position'' that happens to go past
the Earth, could be detected and thereafter ``followed'' in its
receding journey, through its refraction of starlight.
The expected angular shifts were estimated at $\Delta\phi
\approx h$, where $h$ is the gravity wave's strength
when it hits the Earth. Although this effect comes with
some neat features, it could not be
of more than a purely academic interest, since quantitatively
the shifts should be smaller than $10^{-15}$arcsec for
typical sources. Since the best angular resolutions one
can hope for at the present are of the order of $10^{-7}$arcsec
(for radio-waves,) the experimental situation for this effect
seemed as bad as (if not worse than) for previously considered
approaches.

In [22], it was shown how in certain, not uncommon
astronomical configurations,  a similar (though not identical)
refraction effect could be at work, with the difference that
the expected shifts in apparent positions of light sources
are many orders of magnitude larger than in [21].

The desirable configurations here are those that have
{\em the gravity wave
source be situated between (and nearly aligned with) the Earth
and the light source.} The optimal angular shifts are
shown to be
$\Delta\phi\sim \pi h(\Lambda)$. But here $h(\Lambda)$
is the strength
of gravity waves at a distance of one reduced gravitational
wavelength from the source, which is many orders
of magnitude larger than the strength at Earth's level.

It is not too difficult to find astronomical sites, amidst the
rich diversity of the galactic zoo, where these
apparent shifts should be near the limits of observability
(see below and [22-24].)
For example, one can get an extremely good ``alignment''
if one takes as the
gravitational and the electromagnetic sources (hereafter dubbed
SG and SL respectively,) the two components of a binary system (fig.(1)).
Once every orbital period, the light from one member of the
binary (SL) passes very near the other member (SG) before
starting its journey towards the Earth.
Then, the constraint on the alignment becomes  here a constraint on
the inclination with respect to the Earth of the binary's
orbital plane. Given the extreme proximity of the members of
many known tight binaries, this constraint turns out to be
actually satisfied for several known candidates.

In the present paper, we suggest that gravity waves could soon be
detected directly through yet another effect.
The astronomical configurations  exploited here are similar to
those considered in [22], but the effect itself has little
to do (experimentally, of course,) with apparent shifts in angular
positions. What will be calculated here is the Shapiro time delay
experienced by a photon that grazes a gravity-wave source SG,
while on its journey to the Earth. The gravity-wave
induced variations in this time delay will be shown
to fall not far from (if not within) current
time resolution limits. Experimentally, this
latter approach may well be easier to implement than that of [22],
which relied on the more complicated technology of high angular
resolution power.

There is more than one type of astronomical site where one
can hope to detect gravity waves in this fashion. In real life,
several factors have to be weighed in the selection of candidate
sites, including whether the magnitude of the light source SL
is large enough in regard of the available integration time.
More detailed predictions for specific sites will be dealt with
elsewhere [23-24]. Here, we shall first derive the main formulae for the
effect (Section II), and then estimate the orders of
magnitude involved by
focusing on the two specific illustrations of [22].

In the first illustration, the light source
SL and the gravity-wave source SG form a tight binary
system (fig.(1)). SL is a high frequency pulsar (say a millipulsar,)
while SG is a substantially slower neutron star. Once every
binary rotation period, the light pulses from SL pass very near
SG before traveling on towards the Earth. The pulsar's frequency,
as observed from the Earth, should vary on (at least) two different
time scales: one related to the binary's rotation (variation of the
impact parameter of SL's photons with respect to SG,) the other,
the object of our scrutiny, related to the relatively strong
curvature fluctuations radiated by the neutron
star SG.

In the second illustration, SL is again a pulsar, but SG is some
much closer, unrelated gravity-wave source, such a binary of
ordinary  or giant stars, that happens to lie not far from
the direction of the distant pulsar. Because, as shows the
calculation below, the constraint on the alignment of
SG and SL is not too stringent, and because such a large
proportion (close to $50\%$) of the stars in the Galaxy
are in binaries or multiple-star systems,
 the chances of a successful hunt for
good candidates seem fairly high for this second case as well.
Note that we have just come across an earlier reference where
the possibility of using pulsar-timing data to detect gravity waves
from binary stars was envisaged [25].
\clearpage
\centerline{\bf II. Equations for the Effect}
\vspace*{0.5cm}

Take then a spherical transverse-traceless coordinate system
centered on SG. Consider, for simplicity, the effect of only
one of the two polarization components of the gravity waves radiated
by SG. (These two components are expected to be, in most cases,
of comparable strengths.) Consider also that the problem is
 completely
contained in the plane defined by SL,SG and the Earth. This
implies choosing the axes of projection of the wave so that only
deformations of geodesics in that plane are considered.

With these simplifications, a null geodesic stretching from
the light source SL to the Earth, and passing near the
gravitational source SG, can be described by the line-element
\newline\begin{equation}
dt^{2} - dr^{2} - (1+h)r^{2}d\phi^{2} = 0 \  \ .
\end{equation}\newline
$h$ is the gravity-wave amplitude, which can be cast in the form
\newline\begin{equation}
h = {H\over r}\exp \{i\Omega (r-t+t_{ph})\} \ \ ,
\end{equation}\newline
where $H$ is a constant that is characteristic of the source SG.
$\Omega$ is the waves frequency component under consideration.
(A source such as a neutron star could emit gravity waves at
various frequences due to different physical phenomena. See
[1] and references therein.)
$t_{ph}$ fixes the phase of the wave at one end of the geodesic.
Thus, in the following, $t_{ph}$ could be the time that a terrestrial
observer reads on her clock at the arrival of a given photon,
while $t$ is just an internal parameter (e.g. expressible in terms
of $r$ or $\phi$) for the trajectory. The quantity we are
ultimately interested in, the rate of change in the gravity-wave
induced time delay, is the {\em variation, with respect to $t_{ph}$,
of the integral of $t$ from the light source SL to the Earth}.

We have not included the background Schwarzschild curvature in (1).
Usually, the time delay related to that contribution can be easily
subtracted from the observations. However, for extremely close
encounters of SL's photons with SG, (impact parameters smaller than
one gravitational wavelength,) it is not always trivial to disentangle
(dipole) Newtonian contributions from relativistic ones.

Using the radial coordinate $u\equiv 1/r$ we can write (1)
in the form
\newline\begin{equation}
u^{2}t'^{2} - {u'^{2}\over u^{2}} = 1 + h \  \ ,
\end{equation}\newline
where primes indicate derivatives with respect to $\phi$.

For $H=0$, the photon follows a straight trajectory given
by
\newline\begin{equation}
u_{0} = {\sin\phi\over b}  \ \ , \ \ \ \
t_{0} = -b\cot\phi \  \ ,
\end{equation}\newline
where $b$ is the distance of closest approach
(the ``impact parameter'') of the photon from SG's
position in the absence of gravity-waves.

Let us try to solve the problem perturbatively by
writing
\newline\begin{equation}
u = u_{0} + u_{1}  \ \ , \ \ \ \
t = t_{0} + t_{1} \  \ ,
\end{equation}\newline
where the gravity-wave-induced fluctuations
$u_{1}$ and $t_{1}$ should be of first order in $h$.

Using (5), (3) becomes
\newline\begin{equation}
u_{0}^{2}t_{0}'^{2} \left( 1 + 2 {u_{1}\over u_{0}}
 + 2 {t_{1}'\over t_{0}'}  \right)
- {u_{0}'^{2}\over u_{0}^{2}}
\left( 1 + 2 {u_{1}'\over u_{0}'}
 - 2 {u_{1}\over u_{0}}  \right) = 1 + h \  \ .
\end{equation}\newline
Here, as throughout this paper, only first-order
terms are retained in the calculations.

As can be seen from (4), the zeroth order parts of
$u$ and $t$ verify
\newline\begin{equation}
u_{0}^{2}t_{0}'^{2} - {u_{0}'^{2}\over u_{0}^{2}} = 1  \  \ .
\end{equation}\newline
Combining this with (6) yields
\newline\begin{equation}
{t_{1}'\over b} = {h\over 2} -
{1+\cos^{2}\phi\over \sin^{3}\phi} b u_{1}
+ {\cos\phi\over\sin^{2}\phi} b u_{1}'  \  \ ,
\end{equation}\newline
where we used $u_{0}^{2} t_{0}' = 1 \ $ (see (4)).

Hence, the gravity-wave-induced perturbation of a
photon's time of flight from the light source SL
to the Earth, is given by the compact expression
\newline\begin{equation}
\Delta t_{1} \equiv t_{1}(\phi_{final}) - t_{1}(\phi_{initial})
= b^{2} \left[ {\cos\phi\over \sin^{2}\phi} u_{1}
\right]_{\phi_{initial}}^{\phi_{final}} +
{b\over 2} \int_{\phi_{initial}}^{\phi_{final}} h d\phi \  \ .
\end{equation}\newline
Furthermore, the photon paths we are interested in are those
which start and end at the fixed positions of the light
source and the observer, respectively [26]. These paths have
\newline\begin{equation}
u_{1}(\phi_{final}) = u_{1}(\phi_{initial}) = 0 \  \ ,
\end{equation}\newline
and so, the first term in (9) vanishes. (Of course,
this holds only to first
order in $h$. Corrections due, e.g., to the fluctuation
of SL's and the observer's positions are of higher order, and
contribute a negligible amount to the integral.)

We can now write the rate of change in the gravity-wave-induced
time delay (see (2)):
\newline\begin{equation}
\dot{\tau} \equiv
{d\ \over dt_{ph}} \Delta t_{1} =
{i b \Omega\over 2} \int_{\phi_{initial}}^{\phi_{final}} h d\phi  \  \ .
\end{equation}\newline
To first order, we can use (4) to write $h$ as
\newline\begin{equation}
h = {H\over b} \sin\phi
\exp\left\{ ib\Omega{1+\cos\phi\over \sin\phi} \right\}
e^{i\Omega t_{ph}} \  \ .
\end{equation}\newline
Thus, we obtain
\newline\begin{equation}
|\dot{\tau}| \approx {1\over 2} \Omega H \left| e^{i\Omega t_{ph}}
\int_{\phi_{initial}}^{\phi_{final}}d\phi
\sin\phi \exp\left\{ ib\Omega{1+\cos\phi\over \sin\phi} \right\}
\right| \  \ .
\end{equation}\newline
The variation of $|\dot{\tau}|$ with the impact parameter
is plotted in fig.2 for the generic case:
$\phi_{initial} \rightarrow 0 \  , \ \phi_{final}\rightarrow \pi .$

We have neglected here contributions from the
time variation of the impact parameter $b$.
In the solar (Schwarzschild-metric) case,
the variation of $b$ (due to
the Earth's revolution around the Sun,) plays
a central role in making the Shapiro time delay
observable (see e.g. [9]). In contrast, the effect in the
present gravity-wave case is made observable
mainly by the variation of the metric itself.
For most conceivable configurations where the
effect described here might be at work, the
time scale of $b$-variations is much larger
than the gravity-waves typical period (see
next section.)

\clearpage
\centerline{\bf III. Order-of-magnitude predictions}
\vspace*{0.5cm}
The observational consequences of eq.(13) depend
closely on the characteristics of the particular
astronomical
site considered. Here we derive typical orders of
magnitude to be expected from the two types of
astronomical configurations
considered in [22]. A more detailed analysis of a few
actual candidate sites is carried out elsewhere [24].

A word, first, about the additional contribution to
$\dot{\tau}$ of the time variation of the impact
parameter $b$, which we have not included in our
formulae.
As we pointed out earlier, it is that time variation
which makes the time delay observable in the Schwarzschild
case. In the calculation of the solar time delay, for
example, the metric itself is static, so that the only
time dependence in the time-delay formula comes from
the $db/dt_{ph}$ produced by the Earth's revolution around
the Sun. (See (2) for the definition of $t_{ph}$.)
In our radiative case, the contribution to $\dot{\tau}$
from the time variation of the metric far exceeds that
from the time variation of $b$. That is to say,
one usually has $db/dt_{ph} << b\Omega$.
Even when SG and SL form a binary, which is a typical case where
$b$'s time modulation could be expected to play a role,
$db/dt_{ph}$ is of the order of $100$km/sec, or about
$3\times 10^{-4}$ in geometrized units. $b\Omega$, on the other
hand, comes to about $300$ in the same units.

Going back to (13), the integral there is of order unity for
the small impact parameters $b\Omega\approx 1$ (see
fig.(2)),
that is, for $b$ of the order of one reduced gravitational
wavelength $\Lambda\equiv 1/\Omega$. Hence,
\newline\begin{equation}
|\dot{\tau}|_{b\approx \Lambda} \sim \Omega H =
 |h(r=\Lambda)| \  \ .
\end{equation}\newline
This is, of course, several orders of magnitude larger than
similar time-delay effects expected from background
(essentially plane) gravitational waves [29].

For larger values of $b$, the integral in (13) decreases
very  roughly as $1/b\Omega$ (fig.(2)). This implies that
the effect's order of magnitude is given by the very
approximate formula (see (13,14))
\newline\begin{equation}
|\dot{\tau}| \sim {H\over b}
= |h(r=b)|  \  \ .
\end{equation}\newline
Hence, the effect is weaker by
about one order of magnitude when the rays from SL
graze SG at ten, rather than at one
reduced gravitational wavelength.
The observational
significance of this formula will become clear
when we apply it to some known astronomical sites such
as tight binary pulsars (below.)

To put some numbers on the above analysis,
let us consider the same two show-cases of [22].

In the first, the source of gravity waves SG is a binary
star like, say, the binary of giant stars $\mu$-Sco in the
Scorpio constellation. That system should be generating
gravity waves at a frequency of about $1.6\times 10^{-6}$Hz,
and with a strength such that the constant $H$ in (2,12)
is about $6$cm. These numbers produce a maximum time-delay
variation $\dot{\tau}\sim 10^{-15}$ to $10^{-14}$, which
is just about what technology can handle today.
(Note that the gravity-wave period is typically
{\em half} the dynamical time-scale of the source.)

The remaining issue, of course, is how likely or unlikely
it is to find, for a given pulsar (SL), an intervening
binary star (SG) lying sufficiently close to the
direction of that pulsar in the sky. The constraint
is that the angular separation of SL and SG should be
within one order of magnitude of the ratio
$\Lambda/D$, where $D$ is the distance from SG to the
Earth. This means that binaries such as $\mu$-Sco,
which is at $D=109$pc, should be within a few seconds
of arc of the pulsar SL. Note that the reduction in the
probability of finding a good SG candidate for a given SL,
due to the decrease of the
allowed SL-SG angular separation with $D$, is
partially compensated by the fact that many of the known
pulsars lie at very large distances (a few or several kpc.)
Since about half
the stars in the Galaxy are in double or multiple stars,
a solid angle of a few seconds of arc, in most directions,
is not unlikely
to contain a good SG candidate.
The binaries one
encounters in the literature are usually, for obvious reasons, the
closest ones to the Earth (e.g. $\mu$-Sco.)
But, as is clear from all the above, the closeness of
SG to the Earth
is not indispensable for observing this effect.
Thus,
the spectrum of observationally interesting objects is
widely broadened
in this approach. High angular resolution space-based
missions like the Hypparcos satellite (sky mapping with
a resolution of about $10^{-4}$''arc), could be instrumental
in identifying cases of sufficient SL-SG alignment
in the Galaxy.

The second show-case used to illustrate [22] had SL again be
a rapid pulsar, but SG was now some more slowly rotating neutron star,
tied with SL in a binary system (fig.(1)). Take SG to be a neutron
star like Vela, with a frequency of $22$Hz, and a gravity-wave
strength such that $H\sim 10^{-4}$cm, supposing that the
mechanism of gravity-wave generation is something like the
so-called CFS (Chandrasekhar-
Friedman-Schutz) instability [27,28]. Than the maximum time-delay
variation observed in SL's data comes to a few times $10^{-14}$.

The alignment problem is almost automatically solved,
in this case. (We name  below a known
astronomical site
 where sufficient alignment occures naturally.)
We recall that the alignment constraint, for this type
of site, is really a
constraint on the orbital inclination of the SG-SL
binary with respect to the observer's line of sight:
the closer the system is to being an ``eclipsing''
binary, the smaller a minimal impact parameter the light
from SL will reach as SL goes around SG (see fig.(1).)
The numbers used above and in [22] actually allow for orbital
inclinations that are many degrees away from the eclipsing value.
which leaves room for statistically many potential candidates
(see Introduction and below.)

We close this paper with a name and a celestial address of
a site where, according to this study, a truly direct
observation of gravity waves could already be within
the realm of possibility.
Take, for instance, the well documented Taylor pulsar (PSR B1913+16.
See e.g. [30]).
This is a $17$Hz pulsar which is in very close, highly eccentric
orbit around a slightly lighter and much darker neutron
star companion. The gravitational wavelength of the latter should
not be much smaller than about one light-second.
Every $7^{h}45'$ or so, the pulsar comes to within
only {\em half a solar radius} of its companion. That is
also a distance of just about one light-second.
Hence, we have here a  {\em known} case (and there are a few more [24],)
 where $b/\Lambda$ might
be close to unity. It seems therefore safe to predict that,
before long, neutron-star gravity waves
could be
seen directly in this system and alike.
(To give a precise
estimate of $\dot{\tau}$ for this system one
would need, of course, to take into account the precise
disposition
in space of the orbit. But this is unlikely to spoil
 order-of-magnitude estimates.)

In the ongoing search for the effects of {\em background}
gravity-waves in pulsar-timing data, the focus has been,
so far, on stochastic, and in particular on cosmological
gravity waves [29]. Could part of the large data set
already collected bare, hidden, the signature of monochromatic
sources such as neutron stars or binary systems?

\vspace*{2.cm}
\centerline{\bf Acknowledgements}
\vspace*{0.5cm}
I benefited greatly from discussions with
S. Braham, F. Gaitan and W. Shuter. I am particularly indebted
to my teacher W.G. Unruh for his relentless advice and
support, as well as for refusing to buy into earlier
versions of this paper, which lead me to discover some
nasty mistakes therein.

This research was supported, in part,
 by the Cosmology Group in the
Department of Physics, University of British Columbia.

\clearpage
\centerline{\bf Figure captions}
\vspace{2.cm}
\begin{figure}[h]
\vspace{1.cm}
\caption{{\em
A pulsating light source SL (e.g. a centi- or
milli-second pulsar) and a gravity wave source SG
(e.g. a slower neutron star) are locked in a
tight binary system (with a possibly highly
excentric orbit.) Viewed from the Earth,
the two companions are indistinguishable
and SG is invisible for most cases of interest.
But, once every few or several hours, the
frequency of light pulses should be modulated
by SG's gravity waves (see caption of fig.(2).)}}
\end{figure}
\vspace{2.cm}
\begin{figure}[h]
\caption{{\em The magnitude of the effect is shown
to decrease roughly as $1/b$.
This implies that the modulation mentioned in the
caption of fig.(1)
is  of the order of the gravity-wave
amplitude at a distance from the source that is equal
to the impact parameter (i.e. $|\dot{\tau}|\sim h(r=b)$;
see eqs.(17-19).)
Hence, this effect is
many orders of magnitude larger than the effects the
same waves may cause in the vicinity of the Earth.}}
\end{figure}
\end{document}